# Improving the topological charge density operator on the lattice.

C. Christou[a], A. Di Giacomo[b], H. Panagopoulos[a], E. Vicari[b]

[a]Department of Natural Sciences, University of Cyprus, Nicosia.

[b]Dipartimento di Fisica, Università Pisa, and INFN, Sezione di Pisa, Pisa 56100, Italy

We analyze the properties of a class of improved lattice topological charge density operators, constructed by a smearing-like procedure. By optimizing the choice of the parameters introduced in their definition, we find operators having (i) a much better statistical behavior as estimators of the topological charge density on the lattice, i.e. much less noisy; (ii) a multiplicative renormalization much closer to one; (iii) a large suppression of the perturbative tail in the corresponding lattice topological susceptibility.

## 1. Introduction.

In QCD an important role is played by the topological properties. By the axial anomaly, matrix elements or correlation functions involving the topological charge density operator $q(x)$ can be related to relevant quantities of the hadronic phenomenology. We mention the topological susceptibility $\chi$, which is determinant in the solution of the $U(1)_A$ problem, and the on-shell nucleon matrix element of $q(x)$, which can be related to the so-called spin content of the nucleon.

Lattice techniques represent our best source of non-perturbative calculations, however investigating the topological properties of QCD on the lattice is a non-trivial task. In a lattice theory the field is defined on a discretized set and therefore the associated topological properties are strictly trivial. One relies on the fact that the physical continuum topological properties should be recovered in the continuum limit.

Considering a lattice version of $q(x)$, $q_L(x)$, the classical continuum limit must be in general corrected by including a renormalization function. In pure QCD, where $q(x)$ is renormalization group invariant,

$$q_L(x) \rightarrow a^4 Z(g_0^2) q(x) + O(a^6) , \qquad (1)$$

where $Z(g_0^2)$ is a finite function of the bare coupling $g_0^2$, going to one in the limit $g_0^2 \rightarrow 0$. But at $g_0^2 \simeq 1$, where simulations are usually performed, it may be very different from one. The finite renormalization of the widely used lattice operator

$$q_L(x) = -\frac{1}{2^9 \pi^2} \sum_{\mu\nu\rho\sigma=\pm1}^{\pm4} \epsilon_{\mu\nu\rho\sigma} \mathrm{Tr}\left[\Pi_{\mu\nu}\Pi_{\rho\sigma}\right] \qquad (2)$$

($\Pi_{\mu\nu}(x)$ is the product of link variables $U_\mu(x)$ around a $1 \times 1$ plaquette) is indeed very small at $g_0^2 \simeq 1$: for $SU(3)$ $Z(g_0^2 = 1) \simeq 0.18$.

The relation of the zero-momentum correlation of two $q_L$ operators, $\chi_L$, with the topological susceptibility $\chi$ is further complicated by an unphysical background term, which eventually becomes dominant in the continuum limit. Indeed

$$\chi_L(g_0^2) \; = \; a^4 Z(g_0^2)^2 \chi \; + \; M(g_0^2) \; . \qquad (3)$$

Neglecting terms $O(a^6)$, the background term $M(g_0^2)$ can be written in terms of mixings with the unity operator (so-called perturbative tail scaling as $\sim a^0$) and with the trace of the energy-momentum (scaling as $\sim a^4$). In the case of the operator (2) and for $SU(3)$, $M(g_0^2)$ is already dominant at $g_0^2 \simeq 1$ (it is about 85% of $\chi_L$ at $g_0^2 = 1$). As a consequence, using the heating method to evaluate $Z(g_0^2)$ and $M(g_0^2)$ [1], the uncertainty on $\chi$ can be hardly pushed beyond $\simeq 10\%$.

Another problem, which has come up in some studies concerning the lattice determination of the on-shell proton matrix element of $q(x)$ [2], is that the lattice operator (2) is very noisy: a very large statistic and therefore expensive simulations are necessary in order to get a reasonable uncertainty on the final result. In view of a full

QCD lattice calculation the search for a better estimator appears a necessary step.

We study, within the field theoretical approach, the possibility of improving the lattice estimator of $q(x)$ with respect to all the problems listed above, that is we look for local lattice versions of $q(x)$ which are less noisy, have a multiplicative renormalization closer to one, and whose corresponding $\chi_L$ is not dominated by the unphysical backgroung signal $M(g_0^2)$ in the region $g_0^2 \simeq 1$. (Any $\chi_L$ defined from a local $q_L$ will eventually be dominated by its perturbative tail in the continuum limit. To the purpose of evaluate $\chi$ a good result would be to have it small at $g_0^2 \simeq 1$, which should be already in the scaling region.)

## 2. Improved topological charge density operators.

Inspired by the widely used smearing techniques, we consider the following set of operators defined in terms of smeared links $V_\mu^{(i)}(x)$

$$q_L^{(i)}(x) = -\frac{1}{2^9\pi^2} \sum_{\mu\nu\rho\sigma=\pm 1}^{\pm 4} \epsilon_{\mu\nu\rho\sigma} \mathrm{Tr}\left[\Pi_{\mu\nu}^{(i)}\,\Pi_{\rho\sigma}^{(i)}\right], \quad (4)$$

where $\Pi_{\mu\nu}^{(i)}$ is the product of smeared links $V_\mu^{(i)}(x)$ around a $1\times 1$ plaquette. Such smeared links are constructed by the following procedure:

$$\begin{aligned}
V_\mu^{(0)}(x) &\equiv U_\mu(x)\\
\widehat{V}_\mu^{(i)}(x) &= (1-c)V_\mu^{(i-1)}(x) + \frac{c}{6}\sum_{\pm\nu,\nu\neq\mu} S_{\nu\mu}^{(i-1)}(x)\\
S_{\nu\mu}^{(i)}(x) &= V_\nu^{(i)}(x)V_\mu^{(i)}(x+\nu)V_\nu^{(i)}(x+\mu)^\dagger\\
V_\mu^{(i)}(x) &= \frac{\widehat{V}_\mu^{(i)}(x)}{\left[\frac{1}{N}\mathrm{Tr}\,\widehat{V}_\mu^{(i)}(x)^\dagger\,\widehat{V}_\mu^{(i)}(x)\right]^{1/2}} \qquad (5)
\end{aligned}$$

where $V_{-\nu}^{(i)}(x) = V_\nu^{(i)}(x-\nu)^\dagger$. $V_\mu^{(i)}(x)$ and therefore $q_L^{(i)}(x)$ depends on the parameter $c$, which can be tuned to optimize the properties of $q_L^{(i)}$. All these operators have the correct classical continuum limit, i.e. for $a \to 0$ $q_L^{(i)}(x) \to a^4 q(x)$.

Notice that the size of $q_L^{(i)}(x)$ increases with increasing the integer parameter $i$, but they can still be considered as local operators when keeping $i$ fixed while approaching the continuum limit. Anyway, as we shall see that a good improvement with respect to $q_L^{(0)}(x) \equiv q_L(x)$ is already achieved for small values of $i$, by optimizing the choice of the parameter $c$.

The procedure (5) may be used to improve any local operator involving link variables. Smearing methods to improve lattice estimators have been already widely employed in the study of long distance correlations, such as large Wilson loops and hadron source operators.

## 3. Perturbative analysis.

We have calculated $Z(g_0^2)$ to one loop for the once-smeared operator $q_L^{(1)}(x)$ within the Wilson action formulation. We find $Z^{(1)}(g_0^2) = 1 + z_1 g_0^2 + O(g_0^4)$, where

$$\begin{aligned}
\frac{z_1}{N} &= \frac{1}{4N^2} - \frac{1}{8} - \frac{1}{2\pi^2} - 0.15493\\
&+ c\left(0.67789 - \frac{0.24677}{N^2}\right)\\
&+ c^2\left(-0.48436 + \frac{0.03991}{N^2}\right) . \qquad (6)
\end{aligned}$$

At $c = 0$ we recover the non-smeared result: $Z = 1 - 0.535 g_0^2 + O(g_0^4)$ for $N = 2$, $Z = 1 - 0.908 g_0^2 + O(g_0^4)$ for $N = 3$ [3]. As $c$ varies, the following extreme values of $Z$ are obtained: $z_1 = -0.136$ at $c = 0.650$ for $N = 2$; $z_1 = -0.247$ at $c = 0.677$ for $N = 3$. In both cases, the reduction of $z_1$ is quite large, making the one loop estimate more reliable for typical values of $g_0^2$.

For $q_L^{(1)}(x)$, we have also calculated the lowest perturbative contribution to the mixing with the unity operator $P(g_0^2)$, which is a part of the background term $M(g_0^2)$.

$$\begin{aligned}
P(g_0^2) &= g_0^6 \frac{3N(N^2-1)}{128\pi^4} p(c) + O(g_0^8) ,\\
p(c) &= 0.002867 - 0.017685c + 0.048665c^2\\
&- 0.075362c^3 + 0.068526c^4\\
&- 0.034433c^5 + 0.007445c^6 . \qquad (7)
\end{aligned}$$

The minimum of this everywhere-concave polynomial is $p(c = 0.872) = 1.4 \times 10^{-5}$. Thus, for all $N$, the leading order of $P$ diminishes by more than two orders of magnitude compared to its non-smeared value ($c = 0$).

In the presence of dynamical fermions one should take into account the fact that, unlike pure gauge theory, the topological charge density mixes under renormalization with $\partial_\mu j^5_\mu$. The non-renormalizability property of the anomaly in the $\overline{\rm MS}$ scheme means that the anomaly equation should take exactly the same form in terms of bare and renormalized quantities. However the renormalization of $\partial_\mu j^5_\mu(x)$ and $q(x)$ is nontrivial [4]. A renormalization group analysis leads to the following relation valid for all matrix elements of a lattice version $q_L(x)$ of $q(x)$ in the chiral limit [5]:

$$\langle i2N_f q_L\rangle \;=\; Y(g_0^2)\,\langle R\rangle \;, \tag{8}$$

where $Y(g_0^2)$ is a finite function of $g_0^2$, and

$$\langle R\rangle \;\equiv\; \langle\; \partial_\mu j^5_\mu(x)_{R_{\overline{\rm MS}}}\;\rangle \exp \int_{g(\mu)}^{0} \frac{\bar\gamma(\tilde g)}{\beta_{\overline{\rm MS}}(\tilde g)} {\rm d}\tilde g \tag{9}$$

is a renormalization group invariant quantity, where $\partial_\mu j^5_\mu(x)_{R_{\overline{\rm MS}}}$ indicates the operator $\partial_\mu j^5_\mu(x)$ renormalized in the $\overline{\rm MS}$ scheme, and the function $\bar\gamma(g)$ is related to the anomalous dimension of the continuum operators $q(x)$, $\partial_\mu j^5_\mu(x)$ in the $\overline{\rm MS}$ scheme: $\bar\gamma(g) = \frac{1}{16\pi^4}\frac{3c_F}{2}N_f g^4 + O\left(g^6\right)$. Notice that $\langle R\rangle$ is what can be naturally extracted also from experimental data.

In perturbation theory one finds $Y(g_0^2) = 1 + (z_1+y_1)g_0^2 + O(g_0^4)$, where $z_1$ is the coefficient of the $O(g_0^2)$ term of the finite renormalization of $q_L$ in the pure gauge theory (cfr. Eq. (6)), and $y_1$ turns out to be a small number: $y_1 = -0.0486$ for $N=3$ and $N_f = 4$.

## 4. Non-perturbative analysis by the heating method.

Estimates of the multiplicative renormalizations of the operators $q_L^{(i)}(x)$ and of the background term in the corresponding $\chi_L$ can be obtained by using the numerical heating method [6,7], without any recourse to perturbation theory. We applied the heating method to the operators $q_L^{(i)}(x)$ for $i = 1,2$ and for a number of values of $c$ in the region $0 \le c \le 1$. We restricted our analysis to the $SU(2)$ pure gauge theory, expecting no substantial differences for $N = 3$. The measurements were performed at $\beta = 2.6$ ($g_0^2 = 1.5384...$), which is a typical value for the $SU(2)$ simulations with the Wilson action. The estimates of $Z^{(i)}(\beta = 2.6)$ from the plateaus observed when heating an instanton configuration are reported in Table 1, and should be compared with the value $Z(\beta = 2.6) = 0.25(2)$ for the standard operator (2) [8].

Table 1
$Z^{(i)}(\beta = 2.6)$ for $i = 1,2$ and some values of $c$.

| $i$ | $c = 0.6$ | $c = 0.8$ | $c = 1.0$ |
|---|---|---|---|
| 1 | 0.52(3) | 0.57(2) | 0.55(2) |
| 2 | 0.67(2) | 0.74(2) | 0.68(2) |

This analysis confirms the one-loop perturbative calculations, that is the improved operators we considered have a multiplicative renormalization closer to one than that of the initial operator $q_L(x)$. From $Z(\beta = 2.6) \simeq 0.25$ of $q_L(x)$, we pass, optimizing with respect to the parameter $c$, to $Z^{(1)}(\beta = 2.6) \simeq 0.57$ by one improving step, and $Z^{(2)}(\beta = 2.6) \simeq 0.74$ by two improving steps. For larger $i$ we expect to get $Z^{(i)}$ closer and closer to one. On the other hand, we should not forget that increasing the number of improving steps the size of the operator $q_L^{(i)}(x)$ increases, therefore one should find a reasonable compromise taking into account the size of the lattice one can afford in the simulations.

A comparison of the above results for $i = 1$ with the one-loop calculation shows that the contribution of the higher perturbative orders is still non-negligible, but not so relevant as in the case of the operator without improving.

Another important property of the improved operators, we can infer from the heating method results, is that they are much less noisy than $q_L(x)$. A quantititive idea of this fact may come from the quantity $e^{(i)} \equiv \Delta Z^{(i)}/Z^{(i)}$, where $\Delta Z^{(i)}$ is the typical error of the data in the plateau during the heating procedure. We indeed found for $c \simeq 1.0$ and for an equal number of measurements: $e^{(0)}/e^{(1)} \simeq 7$ and $e^{(0)}/e^{(2)} \simeq 18$ .

An estimate of the background signal $M(g_0^2)$ can be obtained by measuring $\chi_L$ on ensembles of configuration constructed by heating the flat configuration [7,8]. The estimates of $M^{(i)}(\beta = 2.6)$ from the plateaus observed in the heating procedure are reported in Table 2, and should be compared with the value $M(\beta = 2.6) = 2.09(5) \times 10^{-5}$ relative to the standard operator (2).

Table 2
$M^{(i)}(\beta = 2.6)$, multiplied by a factor $10^5$, for $i = 1, 2$ and some values of $c$.

| $i$ | $c = 0.6$ | $c = 0.8$ | $c = 1.0$ |
|---|---|---|---|
| 1 | 0.59(2) | 0.36(1) | 0.26(1) |
| 2 | 0.23(1) | 0.13(1) | 0.07(1) |

Notice the strong suppression of the background term in the zero-momentum two-point correlations constructed with the improved operators. For $c \simeq 1$ the reduction is about a factor 8 when performing one improving step, and about 30 by two improving steps. For larger values of $i$, the suppression is expected to be larger and larger.

The suppression of the background term in Eq. (3) together with the relevant increase of $Z$ should drastically change the relative weights of the contributions to $\chi_L$ at $g_0^2 \simeq 1$, i.e. in the relevant region for Monte Carlo simulations. For example if the improvement for $SU(3)$ is similar to that achieved for $SU(2)$, using the optimal operator for $i = 2$ at $g_0^2 = 1$ the unphysical term in Eq. (3) may become a small part of the total, allowing a precise determination of $\chi$ by the field theoretical method.

## 5. Conclusions.

We have analyzed the properties of a class of improved lattice topological charge density operators. Such improved operators look promising for the lattice calculation of the on-shell proton matrix element of the topological charge density operator in full QCD, which is related to the so-called proton spin content. Indeed their use should overcome the difficulty due to the large noise observed in preliminary quenched studies [2], and their multiplicative renormalizations are much closer to one.

They should also provide a relevant improvement in the determination of the topological susceptibility by the field theoretical method in the $SU(3)$ gauge theory, since the unphysical background term should get strongly reduces while the term containing $\chi$ gets enhanced by the larger values of the multiplicative renormalization. This should allow a precise and independent check of the alternative cooling method determinations (see e.g. Ref. [9]), whose systematic errors are not completely controlled. Furthermore the improved operators may open also the road to a more reliable lattice investigation of the behavior of the topological susceptibility at the deconfinement transition, where cooling technique does not seem to give satisfactory results [10].